\documentclass[12pt]{article}
\usepackage{pst-plot,epsf}
\newcommand{\beq}{\begin{equation}}
\newcommand{\eeq}{\end{equation}}
\newcommand{\bea}{\begin{eqnarray}}
\newcommand{\eea}{\end{eqnarray}}
\newcommand{\nobody}{\rule{0ex}{1ex}}

\newcommand{\epm}{e^+e^-}
\newcommand{\emp}{e^-e^+}

\newcommand{\ra}{\rightarrow}

\newcommand{\prog}{\tt ee4f$\gamma$}
\begin{document}
\thispagestyle{empty}
\begin{flushright}
DESY-03-099\\
SFB/CPP-03-16\\
Revised version\\
February 2004\\
\vspace*{1.5cm}
\end{flushright}
\begin{center}
{\LARGE\bf {\tt ee4f$\gamma$}\\[3mm]
 A program for \boldmath{$e^+ e^- \ra 4{\rm f}, 4{\rm f}\gamma$} with
nonzero \\[3mm]
fermion masses\footnote{Work supported
           in part by the Polish State Committee for Scientific Research
           (KBN) under contract No. 2~P03B~045~23, by the European
           Community's Human Potential Program under contract
           HPRN-CT-2000-00149 Physics at Colliders and by DFG
           under Contract SFB/TR 9-03.}}\\
\vspace*{2cm}
Karol Ko\l odziej$^{\rm a,}$\footnote{E-mail: kolodzie@us.edu.pl}
and Fred Jegerlehner$^{\rm b,}$\footnote{E-mail: fjeger@ifh.de}
\vspace{0.5cm}\\
$\nobody^{\rm a}${\small\it
Institute of Physics, University of Silesia, ul. Uniwersytecka 4,
PL-40007 Katowice, Poland}
$\nobody^{\rm b}${\small\it
Deutsches Elektronen-Synchrotron DESY, Platanenallee 6, D-15738 Zeuthen,
Germany}\\
\vspace*{3.5cm}
{\bf Abstract}\\
\end{center}
A computer program {\prog} for calculating cross sections of any four fermion
final state of $e^+ e^-$-annihilation at high energy and the corresponding
bremsstrahlung reaction that is possible in the framework of the
Standard Model is presented. As the fermion masses are arbitrary,
the cross sections for channels that do not contain $e^+$ and/or $e^-$ in
the final state can be computed without any collinear cut,
the on-shell top quark production can be studied and the Higgs boson exchange
can be incorporated in a consistent way.
The program can be used as a Monte Carlo generator of unweighted
events as well.
\vfill
\newpage
{\large \bf PROGRAM SUMMARY}\\[4mm]
{\it Title of program:} {\prog}

{\it Version:} 1.0 (February 2004)

{\it Catalogue identifier:}

{\it Program obtainable from:} CPC Program Library or on request by e-mail
from the authors

{\it Licensing provisions:} none

{\it Computers:} all

{\it Operating systems:} Unix/Linux

{\it Programming language used:} {\tt FORTRAN 90}

{\it CPC Program Library subprograms used:} {\tt RANLUX}
\vspace*{-0.5cm}
\begin{tabbing}
\hspace*{1cm} Cat. Id.  \=  \hspace*{1cm}  RANLUX     \= \hspace*{1cm} 
                                                        79(1994)111 \kill
Cat. Id.    \>     Title     \>              ref. in CPC \\
ACPR        \>    RANLUX     \>    79(1994)111 \\
\end{tabbing}

\vspace*{-0.8cm}

{\it Memory required to execute with typical data:} 4.0 Mb

{\it No. of bits in a word:} 32

{\it No. of bytes in distributed program, including test data, etc.:} 2.016 Mb

{\it Distribution format:} ASCII


{\it Nature of physical problem}\\ Description of all $\epm \ra $ 4
fermions and corresponding bremsstrahlung reactions that are possible
in the Standard Model (SM) to lowest order and with nonzero fermion
masses at center of mass energies typical for next generation linear
colliders.  Such reactions are relevant, typically, for $W$-pair or
intermediate mass Higgs boson production and decay.

{\it Method of solution}\\
Matrix elements are calculated with the helicity amplitude
method. The phase space integration is performed numerically utilizing
a multi-channel Monte Carlo method.

{\it Restrictions on complexity of the problem}\\ No higher order
effects are taken into account, except for assuming the fine structure
constant and the strong coupling at the appropriate scale and partial
summation of those one particle irreducible loop corrections
which are inducing the (fixed) finite widths of unstable particles.

{\it Typical running time}\\
The running time depends strongly on a selected channel and desired precision
of the result.
The results of the appended test run have been obtained on a 800~MHz
Pentium~III processor with the use of Absoft {\tt FORTRAN~90} compiler
in about 9 minutes. In order to obtain a precision level below one per
mille a few million calls to the integrand are required. 
This results in less than one hour running time for the fastest channel
$\epm \ra \nu_{\mu}\bar{\nu}_{\mu}\nu_{\tau}\bar{\nu}_{\tau}$
and more than 100 hours running time for the slowest channel
$\epm \ra \epm\epm $. However, the typical running time for channels
that are relevant for the $W$-pair or Higgs boson production is several hours.



%
{\large \bf LONG WRITE-UP}
\section{Introduction}
Precise tests of the Standard Model (SM) that will become possible at
future high luminosity $\epm$ linear colliders like
TESLA~\cite{Tesla}, the NLC~\cite{Abe:2001wn}, or the
JLC~\cite{Abe:2001gc} require adequate high precision of theoretical
predictions. To meet this requirement is a challenge, as the
complexity of perturbative calculation increases while going to higher
energy and higher luminosity. The spectrum
of possible reactions involves among others the four fermion channels
with possible real photon emission. These reactions, which are of
particular relevance for $W$-pair or intermediate mass Higgs boson
production and decay, receive contributions from several to about one
thousand Feynman diagrams already at the tree level of the SM.

In this article, a technical documentation of a numerical program
{\prog} is presented which allows
for computer simulation of all the four fermion reactions
\beq
\label{born}
e^+(p_1) + e^-(p_2) \ra
f_1(p_3) + \bar{f_2}(p_4) + f_3(p_5) + \bar{f_4}(p_6),
\eeq
and the corresponding bremsstrahlung reactions
\beq
\label{brems}
e^+(p_1) + e^-(p_2) \ra
f_1(p_3) + \bar{f_2}(p_4) + f_3(p_5) + \bar{f_4}(p_6) + \gamma(p_7),
\eeq
where $f_i$, $i=1,2,3,4$ make up a four fermion final state possible
in the SM and the particle momenta have been indicated in
parentheses.  The physics contents of the program has been discussed
in earlier works \cite{JK1}--\cite{JK3}, where the details on the
underlying physics, the method of calculation and physical results
obtained with {\prog} have been presented.

In the approximation of massless fermions, all possible four fermion
channels for $\epm \ra 4{\rm f}$ have been investigated
in~\cite{Berends:1995xn} ({\tt EXCALIBUR}) and those for $\epm \ra
4{\rm f}\gamma$ in~\cite{Denner:1999gp} ({\tt RacoonWW}). {\prog} makes
possible computations with nonzero fermion
masses.  Keeping nonzero fermion masses will be important in cases
when predictions at the 1\% accuracy level are
required~\cite{JK1}--\cite{JK3}. Finite masses
also provide a natural regularization of distributions which become
singular in the massless limit. Massive calculations thus provide
reliable benchmarks for massless calculations with cuts. The latter
are much simpler and hence much faster than calculations with massive
codes.

{\prog} may be considered as a building block (the soft plus
hard bremsstrahlung part) of a complete $O(\alpha)$ calculation of
the processes $\epm \ra 4{\rm f}$. Such calculations have been
attempted in~\cite{Vicini:1998iy} (see also
\cite{Aeppli:1991zp}) and have been completed in the double pole approximation
for $W$-pair production \cite{Denner:2000bj} ({\tt RacoonWW}) and
\cite{Placzek:2001ng} ({\tt KORALW/YFSWW}) (see also
\cite{Fleischer:1995sq} ({\tt EEWW})). The double pole approximation
incorporates the one-loop corrections for on-shell
$W$-pair production~\cite{rc} and the subsequent decay of $W$'s into
fermion-pairs~\cite{decay}.

Several existing programs, some of them being general purpose
packages, may be utilized for tree-level calculations of
$\epm \ra 4{\rm f},\;4{\rm f}\gamma$ with massive fermions.
These are {\tt GRACE/BASES}~\cite{Tanaka:1991wn}, {\tt
MADGRAPH/HELAS}~\cite{Stelzer:1994ta}, {\tt
CompHEP}~\cite{Boos:1994xb} (squared matrix element technique), {\tt
WPHACT}~\cite{Accomando:1997es}, {\tt
NEXT\-CALI\-BUR}~\cite{Berends:2000gj} (initial state radiation
photons generated via the structure function approach), {\tt
HELAC/PHEGAS}~\cite{Kanaki:2000ey} (recursive Dyson-Schwinger equation
approach), {\tt WRAP}~\cite{Montagna:2001uk} ({\tt
ALPHA}~\cite{Caravaglios:1995cd} algorithm), {\tt
O'Mega/WHIZARD}~\cite{Moretti:2001zz} and {\tt AMEGIC++}~\cite{Krauss}.
Most of the codes work on the
basis of helicity amplitudes and some use the structure function approach
to generate the photons. Except for {\tt NEXT\-CALI\-BUR}, which is
specialized to $\epm \ra 4{\rm f}$, all other programs allow to
generate and evaluate amplitudes for other type of processes. For
more details and comparisons we refer to the {\it Four Fermion Working
Group Report}~\cite{Grunewald:2000ju}. 

Results obtained with {\prog} reproduce nicely, within one standard deviation,
the results for all channels of (\ref{born}) and (\ref{brems}) in 
the massless fermion limit of SM that have been surveyed in Table~1 of
\cite{Denner:1999gp}. Moreover, results on (\ref{born}) 
with nonzero fermion masses of \cite{JK3} that had been obtained
with {\prog} have been reproduced with 
{\tt AMEGIC++}~\cite{Krauss1}, except for a discrepancy for the rather 
unphysical channels 
containing on-shell $t\bar{t}$-quark-pair, which
could be traced back to a different implementation of the top
quark width in the propagator\footnote{The misprinted entry for $\sigma$ of
$\epm \ra t\bar{b}b\bar{t}$ at $\sqrt{s}=500$~GeV in Table~1 of \cite{JK3}
should be replaced with 1.2753(7).}. 

The main advantage of the present program is that it allows for
relatively fast computation cross sections of both (\ref{born})
and (\ref{brems}). As fermion masses are not neglected, the cross sections
for channels that do not contain $e^+$ and/or $e^-$ in
the final state can be calculated without any kinematical cuts. A number of
options have been implemented in the program which make possible calculation
of the cross sections while switching on and off different subsets of the
Feynman diagrams.
Besides of a possibility of taking into account solely the electroweak
contributions, or switching off the Higgs boson exchange diagrams,
the program allows also for a simplified treatment of reactions
(\ref{born}), (\ref{brems})
by utilizing  two different narrow width approximations:
for $W$ bosons
\bea
\label{doubleW}
  e^+(p_1) + e^-(p_2) & \ra &
      {W^+}(p_{34}) + {W^-}(p_{56}) \\
&\ra &
\label{final1}
f_1(p_3) + \bar{f_2}(p_4) + f_3(p_5) + \bar{f_4}(p_6),
\eea
and for a $\bar{t}$-quark
\bea
\label{eett}
  e^+(p_1) + e^-(p_2) & \ra &
      t(p_{3}) + \bar{t}(p_{456})  \\
\label{final2}
&\ra &
t(p_3) + \bar{b}(p_4) + f_3(p_5) + \bar{f_4}(p_6).
\eea

\section{Scheme of the calculation}
The necessary matrix elements of reactions (\ref{born}) and
(\ref{brems}) are calculated with the helicity amplitude method described
in \cite{KZ} and \cite{JK1} and phase space integrations are performed with
the Monte Carlo (MC) method.

In order to improve the
convergence of the MC integration the most relevant peaks of
the matrix element squared related to the Breit-Wigner shape of the
$W, Z$, Higgs
and top quark resonances as well as to the exchange of a massless photon or
gluon have to be mapped away. 
As it is not possible to find out a single parametrization of the
multi-dimensional phase space which would allow to cover the whole resonance
structure of the integrand, it is necessary to utilize a multi-channel
MC approach \cite{MC}. The reader is referred to \cite{tt6f} for
more details on the multi-channel MC algorithm, on which the MC integration
and event generation is based. In the next two subsections we collect
phase space parametrizations and mappings of integration 
variables that are used in the program.

\subsection{Phase space parametrizations}

The basic phase space parametrizations which are used in the program are
described in the following. The 4 particle phase space of reaction (\ref{born}) is
parametrized in 2 different ways:
\bea
\label{dpsb1}
{\rm d}^7Lips &=& (2\pi)^{-7} \;
  \frac{\left|\vec{p_3}+\vec{p_4}\right|}{4\sqrt{s}}\;{\rm d}\cos\theta_{34}
        {\rm d}s_{34} {\rm d}s_{56}
        {\rm d} PS_2\left(s_{34},m_3^2,m_4^2\right)
        {\rm d} PS_2\left(s_{56},m_5^2,m_6^2\right),\\
\label{dpsb2}
{\rm d}^7Lips &=& (2\pi)^{-7} \;
  \frac{\left|\vec{p_3}\right|}{4\sqrt{s}}\;{\rm d}\cos\theta_{3}
        {\rm d}s_{456} {\rm d}s_{56}
        {\rm d} PS_2\left(s_{456},s_{45},m_6^2\right)
        {\rm d} PS_2\left(s_{56},m_5^2,m_6^2\right),
\eea
where $s=(p_1 + p_2)^2$, $s_{ij...}=(p_i + p_j + ...)^2$
and $ {\rm d} PS_2(p^2,q^2,r^2)$
is a two particle (subsystem) phase space element defined by
\bea
\label{dps}
 {\rm d} PS_2\left(p^2,q^2,r^2\right) = \delta^4\left( p - q - r \right) \;
   \frac{{\rm d}^3q}{2q^0} \; \frac{{\rm d}^3r}{2r^0}
    = \frac{|\vec{q}|}{4\sqrt{p^2}} {\rm d} \Omega,
\eea
with $\vec{q}$ being the momentum and $\Omega$ the solid angle of one
of the particles (subsystems) in the relative center of mass system,
$\vec{q} + \vec{r} = 0$. Momenta $\vec{p_3} + \vec{p_4}$, $\vec{p_3}$
and the corresponding polar angles $\theta_{34}$, $\theta_{3}$ in
Eqs.~(\ref{dpsb1}) and (\ref{dpsb2}) are defined in the centre of mass
system (c.m.s.) frame
with a positive direction of the $z$ axis determined by
$\vec{p_1}$. Using the rotational symmetry with respect to the
$e^+e^-$ beam axis, the azimuthal angles of $\vec{p_3}+\vec{p_4}$ in
Eq.~(\ref{dpsb1}) and $\vec{p_3}$ in Eq.~(\ref{dpsb2}) are set to zero
which reduces the dimension of the phase space integral from 8 to
7. Parametrization (\ref{dpsb1}) covers the kinematical situation
where the final state fermions make up two 2-particle subsystems of
approximately equal invariant mass. Similarly, the parametrization
(\ref{dpsb2}) corresponds to the production of a 1- and 3-particle
subsystem with the latter decaying again into 1- and 2-particle
subsystems.

The 5 particle phase space of reaction (\ref{brems}) is
parametrized in 3 different ways. Parametrizations
\bea
\label{dps1}
{\rm d}^{10}Lips&=&(2\pi)^{-10} \; \frac{E_7}{2} \;{\rm d}E_7
{\rm d}\cos\theta_7 \; {\rm d} PS_2\left(s',s_{34},s_{56}\right)
{\rm d}s_{34} {\rm d}s_{56}   \nonumber \\ &&~~\times ~~
        {\rm d} PS_2\left(s_{34},m_3^2,m_4^2\right)
        {\rm d} PS_2\left(s_{56},m_5^2,m_6^2\right),\\
\label{dps2}
{\rm d}^{10}Lips&=&(2\pi)^{-10} \; \frac{E_7}{2} \;
   {\rm d}E_7 \; {\rm d}\cos\theta_{7}
        {\rm d} PS_2\left(s',s_{456},m_3^2\right)
        {\rm d}s_{456} {\rm d}s_{56}   \nonumber \\ &&~~\times ~~
        {\rm d} PS_2\left(s_{456},s_{56},m_4^2\right)
        {\rm d} PS_2\left(s_{56},m_5^2,m_6^2\right),
\eea
with $s'=(p_1 + p_2 - p_7)^2$ and the photon energy $E_7$
and the solid angle $\Omega_7$, being defined in the c.m.s., cover the initial
state photon emission.
The final state photon emission is covered by the following
phase space parametrization
\bea
\label{dps3}
{\rm d}^{10}Lips &=& (2\pi)^{-10}
\frac {\lambda^{1/2}(s,s_{347},s_{56})}{8s}\;\nonumber \\
&\times& \frac{1}{8} {\rm d}s_{347} {\rm d}s_{56}{\rm d}\cos\theta_{347}
{\rm d}E_3 {\rm d}E_7 {\rm d}\Omega_3 {\rm d}\phi_{37}
        {\rm d} PS_2\left(s_{56},m_5^2,m_6^2\right),
\eea
where the polar angle $\theta_{347}$ of the momentum
${\bf p}_3+{\bf p}_4+{\bf p}_7$ is defined in the c.m.s.;
the energy $E_3$ and spherical angle $\Omega_3$ of the final state
particle that radiates a photon, the photon energy $E_7$ and the azimuthal
angle $\phi_{37}$ with respect to the momentum of the radiating particle
are defined in the frame where ${\bf p}_3+{\bf p}_4+{\bf p}_7=0$.
Parametrizations (\ref{dps1}--\ref{dps3}) are also used with different
permutations of the external particle momenta and we have made use of
the rotational symmetry with respect to the beam axis in order
to reduce the dimension of integration from 11 to 10.

The invariant masses $s_{ijk...}$ in Eqs.~(\ref{dpsb1})--(\ref{dps3})
should, if possible, correspond to the virtuality of the propagators
of the gauge bosons, Higgs boson and/or top quarks.  The peaks related
to the propagators are then smoothed out by performing suitable
mappings
\cite{tt6f}.

\subsection{Collinear, soft and quasi real photons}

This subsection is dedicated to the issue of
collinear, soft and quasi real photons that is of particular 
relevance for the precision of phase space integration which
is performed in the program.

The strong collinear peaking behaviour of the squared matrix element 
of reaction (\ref{brems}) corresponding to the radiation off
the initial state positron 
\beq
\label{col1}
1/\left[(p_1-p_7)^2-m_e^2\right]
=\;\sim 1/\left[E_7 (1 - \beta\cos\theta_7)\right]
\eeq 
is eliminated by the mapping
\bea
\cos\theta_7=\frac{1}{\beta_e}\left[1-(1+\beta_e)/r_e^x\right],
\eea
with $r_e = (1+\beta_e)/(1-\beta_e)$ and $\beta_e=\sqrt{1-4m_e^2/s}$ 
being the velocity of the electron in the c.m.s. Similarly,
the collinear peaking related to the radiation off the initial state electron
\beq
\label{col2}
1/\left[(p_2-p_7)^2-m_e^2\right]=
\;\sim 1/\left[E_7 (1 + \beta\cos\theta_7)\right]
\eeq
is dealt with the mapping
\bea
\cos\theta_7=\frac{1}{\beta_e}\left((1-\beta_e)r_e^x-1\right).
\eea
The $\sim 1/E_7$ peaking of the bremsstrahlung photon spectrum 
of Eqs.~(\ref{col1}) and (\ref{col2}) is eliminated by the mapping
\bea
\label{mape7}
E_7=E_7^{\rm min}\left(E_7^{\rm max}/E_7^{\rm min}\right)^{x},
\eea
where $E_7^{\rm min}$ and $E_7^{\rm max}$ are the lower and upper limit of the
photon energy $E_7$.

The collinear and soft photon peaking corresponding to
radiation off a final state fermion, e.g. $f_2(p_4)$, of reaction 
(\ref{brems}) 
\beq
\label{colf}
1/\left[(p_4+p_7)^2-m_4^2\right] \sim
1/(p_4 \cdot p_7) \sim 1/(C_3-E_3)
\eeq
is mapped away with the following mapping
\bea
\label{mape3}
E_3=C_3-(C_3-m_3)\left(\frac{C_3-E_3^{\rm max}}{C_3-m_3}\right)^x,
\eea
where $C_3=\sqrt{s_{347}}/2+(m_3^2-m_4^2)/(2\sqrt{s_{347}})$,
with $s_{347}=\left(p_3+p_4+p_7\right)^2$.
In Eqs.~(\ref{col1}--\ref{mape3}) and in Eqs.~(\ref{cosff1})--
(\ref{cosp}) below, $x$ denotes a random variable uniformly
distributed in the interval $[0, 1]$.

In the soft photon limit, the integration over the photon phase
space can be performed analytically
\bea
\label{softint}
\left|{\rm d}\sigma_{\gamma}\right|_{|{\bf p}_7| < \omega}\!\!&=&\!\!
-\frac{1}{(2\pi)^3}\int_{|{\bf p}_7| < \omega}
\frac{{\rm d}^3p_7}{2E_7}
\left(  g_{\gamma 1} {{p_1} \over {p_1 \cdot p_7}}
       - g_{\gamma 2} {{p_2} \over {p_2 \cdot p_7}}
       + g_{\gamma 3} {{p_3} \over {p_3 \cdot p_7}}
       - g_{\gamma 4} {{p_4} \over {p_4 \cdot p_7}}\right.\nonumber\\
  & & \qquad\qquad\qquad\qquad
 + \left.  g_{\gamma 5} {{p_5} \over {p_5 \cdot p_7}}
       - g_{\gamma 6} {{p_6} \over {p_6 \cdot p_7}}\right)^2
         {\rm d}\sigma_0= 
-\sum_{i,j=1}^5 g_{\gamma i}g_{\gamma j} I_{ij}^{\omega},
\eea
where $g_{\gamma i}$ denote the SM couplings of the photon to $i$-th
fermion flavour.  The bremsstrahlung integrals $I_{ij}^{\omega}$,
which are defined by
\bea
\label{iomega}
I_{ij}^{\omega}=\frac{1}{(2\pi)^3}\int_{|{\bf p}_7| < \omega}
\frac{{\rm d}^3p_7}{2E_7}\frac{p_i\cdot p_j}{(p_i\cdot p_7)(p_j\cdot p_7)}\;\;,
\eea
for $i \neq j$ may be found in Section 7 of Ref.~\cite{tHV}. For 
$i=j$ we have
\bea
\label{iomegaii}
I_{ii}^{\omega}=\ln\frac{2\omega}{m_{\gamma}}-\frac{1}{\beta_i}
\ln\frac{1+\beta_i}{1-\beta_i},
\eea
where $\beta_i$ is the velocity of the radiating particle in the c.m.s.
and $m_{\gamma}$ denotes a fictitious mass of the photon that has
been introduced in order to regularize the infrared divergence. Splitting up
bremsstrahlung reaction (\ref{brems}) into the hard and soft photon  parts
may serve as a test of the quality of integration over the real photon phase
space. The inclusive bremsstrahlung cross section including both the soft 
and hard photon parts should be independent of the soft photon energy 
cut $\omega$.
This independence has been successfully tested for different channels
of (\ref{brems}) in \cite{JK1}--\cite{JK3}. The user may repeat such
tests on his own by running the program with {\tt isoft = 1}
and {\tt ihard = 1} with different values of {\tt ecut = $\omega$}. 
See Section~3 for explanation of the notation.

Particular attention has also been paid to the channels of (\ref{born}) 
and (\ref{brems}) that include a light fermion pair, or an electron and/or 
a positron in the final state. Matrix elements of such
reactions develop semisingularities at phase space regions where
the virtuality of the intermediate photon approaches the squared
invariant mass of the light fermion pair, or 
if the photon exchanged in the $t$-channel becomes quasi-real, as may
happen in case of  the final state electron and/or positron being hardly
scattered.

The decay of the intermediate photon into a light fermion-antifermion pair
of invariant mass $s_{f\bar{f}}=p_{f\bar{f}}^2=(p_f+p_{\bar{f}})^2$ 
is accounted for by the following mapping of the corresponding 
$\sim 1/s_{f\bar{f}}$ pole in the Feynman propagator
\beq
 s_{f\bar{f}}=s_{f\bar{f}}^{\rm min} 
  \left(s_{f\bar{f}}^{\rm max}/s_{f\bar{f}}^{\rm min}\right)^x,
\eeq
with $s_{f\bar{f}}^{\rm min} = (m_f + m_{\bar{f}})^2$
and $s_{f\bar{f}}^{\rm max}$ equal to the kinematical upper limit
that is computed in the program for any specific phase space parametrization.
If this virtual photon becomes quasi-real it is preferably emitted
in the direction parallel to initial beams. 
The cosine of its angle with respect to the beam is then generated 
with
\beq
\label{cosff1}
       \cos\theta = y-(y+1) r^x,
\eeq
in case it is parallel to the $e^+$ beam, or according to the formula
\beq
\label{cosff2}
       \cos\theta = (y-1)r^{-x} - y,
\eeq
if it is parallel to the $e^-$ beam. In Eqs.~(\ref{cosff1}) and
(\ref{cosff2}),
\beq
y=\frac{\sqrt{s'} E_{f\bar{f}}-s_{f\bar{f}}}
       {\sqrt{s'} \left|\vec{p}_{f\bar{f}}\right|},
\qquad {\rm and} \qquad r=\frac{y-1}{y+1},
\eeq
with $E_{f\bar{f}}=(s_{f\bar{f}}+|\vec{p}_{f\bar{f}}|^2)^{1/2}$; 
$s'=s$ for reaction (\ref{born}) and 
$s'=\left(p_1+p_2-p_7\right)^2$ for reaction (\ref{brems}). 

The pole corresponding to the exchange of a $t$-channel photon
is mapped away by generating cosine of the final state electron angle 
$\theta_{e^-}$ according to 
\beq
\label{cosm}
       \cos\theta_{e^-}=(y-1) r^x  -y,
\eeq
and cosine of the final state positron angle 
$\theta_{e^-}$ with
\beq
\label{cosp}
       \cos\theta_{e^+}=y-(y+1)/r^x.
\eeq
In Eqs.~(\ref{cosm}) and (\ref{cosp}), $r=(y+1)/(y-1)$ and
$y=(E_i E_f-m_e^2)/(|\vec{p}_i||\vec{p}_f|)$, with
$E_i$ and $\vec{p}_i$ ($E_f$ and $\vec{p}_f$) being the energy and momentum
of $e^-$ or $e^+$ in the initial (final) state.

For the charged current reactions, the integration
over $\cos\theta_{e^{\pm}}$ can be performed to the very kinematical
limit corresponding to
\beq
t_0=-m_e^2\left(E_i-E_f\right)^2/\left(E_iE_f\right).
\eeq
However, for some neutral current channels of reactions (\ref{born}), 
(\ref{brems}), in particular those including another light fermion-antifermion
pair in addition to $\epm$ in the final state, it may lead to a quasi-real 
pole $\sim 1/t$ in the photon propagator, corresponding to 
$e^+$ or $e^-$ passing from the ininitial to final state with almost
no interaction. This singularity can be avoided by imposing a cut
on the angle of the final state electron or positron with respect to
the beam.

\section{Description of the program}
{\tt ee4f$\gamma$} is a package written in {\tt FORTRAN 90}. It consists of
93 files including a makefile. They are stored in one working directory.
The user should
specify the physical input parameters in module {\tt inprms.f} and select
a number of options in the main program {\tt ee4fg.f}.

\subsection{Program input}
The default values of the input parameters and options used in the program
are those specified below.
\subsubsection{Physical parameters}
The initial physical parameters to be specified in module {\tt inprms.f}
are the following.

Boson masses and widths (GeV):
\begin{itemize}
\item {\tt  mw=80.423} GeV, the W mass,
\item {\tt  gamw=2.118} GeV, the W width,
\item {\tt  mz=91.1876} GeV, the Z mass,
\item {\tt  gamz=2.4952} GeV, the Z width,
\item {\tt  mh=115.0} GeV, the Higgs mass,
\item {\tt  gamh=0} GeV, the Higgs width.
      If {\tt gamh = 0}, then the Higgs width is calculated according to
      the lowest order prediction of the SM.
\end{itemize}
The electroweak (EW) mixing parameter {\tt sw2} is then calculated from\\[2mm]
\centerline{\tt sw2 = 1-mw2/mz2 }
with
\begin{itemize}
\item {\tt mw2 = mw**2} and {\tt mz2 = mz**2} in the fixed width scheme,
\item {\tt mw2 = mw**2 - i*mw*gamw} and {\tt mz2 = mz**2 - i*mz*gamz}
      in the complex mass scheme.
\end{itemize}
Coupling constants:
\begin{itemize}
\item {\tt ralp0 = 137.03599976,} the inverse of the fine structure constant in
      the Thomson limit,
\item {\tt gmu=1.16639}$\times {\tt 10}^{\tt -5}$ GeV$^{-2}$, the Fermi
      coupling constant,
\item {\tt  alphas=0.1172}, the strong coupling constant at scale {\tt mz}.
\end{itemize}
The inverse of the fine structure constant at scale {\tt mw}, {\tt ralpw}, is
then calculated from\\[2mm]
\centerline{\tt  ralpw=4.44288293815837/(2*sw2*gmu*mw**2).}

Fermion masses and widths:
\begin{itemize}
\item {\tt me = 0.510998902} MeV, {\tt game = 0} MeV, for the electron,
\item {\tt mmu = 105.658357} MeV, {\tt gammu = 0} MeV, for the muon,
\item {\tt mtau=1.77699} GeV, {\tt gamtau = 0} GeV, for a tau lepton,
\item {\tt mu = 5} MeV, {\tt gamu = 0} MeV, for an up quark
\item {\tt md=9} MeV, {\tt gamd = 0} MeV, for a down quark,
\item {\tt mc=1.3} GeV, {\tt gamc = 0} GeV, for a charm quark,
\item {\tt ms=150} MeV, {\tt gams = 0} MeV, for a strange quark,
\item {\tt mt=174.3} GeV, {\tt gamt = 1.5} GeV, for a top quark,
\item {\tt mb=4.4} GeV, {\tt gamb = 0} GeV, for a bottom quark.
\end{itemize}
No. of colors and conversion constant:
\begin{itemize}
\item {\tt  ncol=3}, the number of colors,
\item {\tt  convc=0.389379292}$\times 10^{12}$ fb ${\rm GeV}^2$, a
      conversion constant.
\end{itemize}
\subsubsection{Main options}
The following main options should be selected in the main program
{\tt ee4fg.f}.
\begin{itemize}
\item The number of different center of mass (CMS) energies {\tt ne}\\
{\tt ne = 1}. {\em Recommended if unweighted events are to be
generated}.
\item The actual values of the CMS energies in the array {\tt aecm} of size
{\tt ne}:\\
{\tt aecm=(/500.d0/)}.
\item The final state of (\ref{born}) by selecting a value of {\tt iproc}
corresponding to a specific channel from the list contained in the file,
{\em e.g.} \\
{\tt iproc=1}\\
corresponds to $\epm \ra  u \bar{d} \mu^- \bar{\nu_{\mu}}$.
The bremsstrahlung is then calculated for the same channel.
\item Whether or not to calculate the Born cross section,
{\tt iborn = 1 (yes) / else (no)}, with {\tt ncall0}
calls to the integrand\\
{\tt iborn = 1}\\
{\tt ncall0 = 100000}. {\em Recommended No of calls is a few millions.}
\item Scan the Born cross section with {\tt nscan0} calls, {\tt iscan0 =
1(yes)/else(no)}, in order to find the dominant kinematical channels, adjust
integration weights and find out the maximum value of the cross section\\
{\tt  iscan0 = 1}.   {\em This option is strongly recommended.}\\
{\tt  nscan0 = 2000}. {\em A few thousands is recommended.}
\item Whether or not to calculate the soft bremsstrahlung cross section,
{\tt isoft = 1 (yes) / else (no)}, with {\tt ncalls}
calls to the integrand\\
{\tt isoft = 1}\\
{\tt ncalls = 100000}. {\em Recommended No of calls is a few millions.}
The integration weights are the same as those used for the Born cross
section.
\item Whether or not to calculate the hard bremsstrahlung cross section,
{\tt ihard = 1 (yes) / else (no)}, with {\tt ncallh}
calls to the integrand\\
{\tt ihard = 1}\\
{\tt ncallh = 100000}. {\em Recommended No of calls is a few millions.}
\item Scan the hard bremsstrahlung cross section with {\tt nscanh} calls,
{\tt iscanh = 1(yes)/else(no)}\\
{\tt  iscanh = 1}.   {\em This option is strongly recommended.}\\
{\tt  nscanh = 2000}. {\em A few thousands is recommended.}
\item Generate the unweighted events or not, {\tt imc = 1(yes)/else(no)}?\\
{\tt  imc = 0.}\\
No standard event record is used. If {\tt imc = 1}, then the final state
particle momenta of the accepted unweighted events are printed in
file {\tt events.dat}.
\item Include the Higgs boson exchange, {\tt ihiggs = 1(yes)/else(no)?}\\
{\tt ihiggs = 1}
\item Calculate the electroweak contributions only, {\tt iew =
1(yes)/else(electroweak  + QCD \\ Feynman diagrams)?}\\
{\tt iew = 0}.
\item Whether or not to calculate approximated cross section of (\ref{born})
{\tt itopa = 1 (yes) / else (no)}, with {\tt ncalla}
calls to the integrand, relevant only for top quark pair production
with an on-shell top quark\\
{\tt itopa = 0}\\
{\tt ncalla = 20000}. {\em Recommended No of calls is a few hundred
thousands.}
\item Whether or not to calculate approximated cross section of (\ref{doubleW})
{\tt iwwa = 1 (yes) / else (no)}, relevant only for $W$-pair production\\
{\tt iwwa = 0}\\
This uses an analytic formula and does not require MC integration.
\item Choose the scheme: {\tt ischeme = 1(complex mass scheme)/else(fixed
width scheme)}\\
{\tt  ischeme = 0}
\item If {\tt ischeme=1}, then should {\tt alpha\_W} be complex
{\tt (iaplw=1)} or real {\tt (ialpw=0)}?\\
{\tt ialpw=0.}
\end{itemize}
\subsubsection{Auxiliary options}
\begin{itemize}
\item Choose gauge: {\tt iarbg = 0(unitary gauge)/1(arbitrary linear gauge)},
      relevant only for the nonradiative top quark production\\
      {\tt  iarbg = 0}
\item If {\tt iarbg=1}, then choose gauge parameters ($\xi \ge 10^{16}$
      corresponds to the unitary gauge):\\
      {\tt ksia=1.d0}\\
      {\tt ksiz=1.d16}\\
      {\tt ksiw=1.d16}
\item Specify the soft photon energy cut:\\
      {\tt ecut=1.d0}
\item Impose cuts, {\tt icuts > 0(yes)/else(no)?}\\
      {\tt  icuts = 1}
\item If {\tt  icuts = 1}, then specify the kinematical cuts:\\
  \begin{itemize}
  \item {\tt  ctlb = 0.985} -- cosine of the charged lepton-beam
                                              angle,
  \item {\tt  ctll = 1} -- cosine of the angle between charged leptons,
  \item {\tt  ctlq = 1} -- cosine of the charged lepton-quark angle,
  \item {\tt  ctgb = 0.985} -- cosine of the photon-beam angle,
  \item {\tt  ctgl = cos(5./180*pi)} -- cosine of the photon-charged lepton
                                                        angle,
  \item {\tt  ctgq = cos(5./180*pi)} -- cosine of the photon-quark angle,
  \item {\tt  ecutg = ecut} GeV -- minimum hard photon energy,
  \item {\tt  ecutl = 5} GeV -- minimum charged lepton energy,
  \item {\tt  ecutq = 0} GeV -- minimum quark energy,
  \item {\tt  mqq = 10} GeV -- minimum invariant mass of a quark pair,
  \item {\tt  mll = 0} GeV -- minimum invariant mass of a charged lepton pair,
  \end{itemize}
\item If {\tt  icuts > 1}, then specify the kinematical cuts on a final
      state electron or positron:
  \begin{itemize}
  \item {\tt theb=5} -- minimum angle of a final state electron and/or
        positron with respect to the beam.
  \end{itemize}
\item Calculate distributions, {\tt idist = 1(yes)/0(no)?}\\
{\tt  idis=0}
\item If {\tt idis = 1}, then specify parameters of the distributions:\\
      {\tt   nbs = (/ n1, ..., n8/)} -- numbers of bins in each distribution.\\
      The corresponding lower and upper bounds of the distribution arguments\\
      {\tt   xmin = (/ xl1, ..., xl8/)} -- lower bounds,\\
      {\tt   xmax = (/ xu1, ..., xu8/)} -- upper bounds,\\
      must be specified below in the same file.
      Constants {\tt xli, xui} should be of type {\tt real(8)} and {\tt ni}
      of type {\tt integer}. The number of desired distributions and the
      maximum number of bins, {\tt nbmax = max\{n1, ..., n8\}} should be
      specified in a module {\tt distribs.f}.
\item The maximum of the fully differential cross section {\tt crmax},
      relevant only if {\tt iscan0 = 0} or {\tt iscanh = 0},\\
      {\tt  crmax=1000}.
\end{itemize}

\subsection{Routines of {\tt ee4fg}}

The main program {\tt ee4fg}, each subroutine, function or module
are located in a file named the same way as the routine itself, except
for subroutines {\tt eev, eve} and {\tt vee}, contained in {\tt
gwwv.f}, and functions: {\tt srr, src} and {\tt scc}, contained
in {\tt dotprod.f}. The package consists of the following routines.

\begin{itemize}
\item {\tt SUBROUTINE} {\tt boost} -- returns a four vector boosted
      to the Lorentz frame of velocity {\bf --v}.
\item {\tt SUBROUTINE} {\tt cancuts} -- checks if generated particle
      momenta satisfy kinematical cuts.
\item {\tt SUBROUTINE} {\tt carlos} -- the MC integration routine.
\item {\tt SUBROUTINE} {\tt couplsma} -- returns SM couplings.
\item {\tt FUNCTION} {\tt cross} -- calculates
      cross sections of (\ref{born}) and (\ref{brems}).
\item {\tt FUNCTION} {\tt crosstopa} -- calculates approximated cross
      section of (\ref{born}) for top quark pair production
      with an on-shell top quark in the narrow anti-top width approximation.
\item The {\tt MAIN PROGRAM} {\tt ee4fg}.
\item {\tt SUBROUTINE} {\tt cseeww} -- returns the total unpolarized
      cross section of on-shell $W$-pair production.
\item {\tt SUBROUTINE} {\tt cutscrit} -- specifies the class of a process
      (\ref{born}), {\em i.e.}, whether the process is purely leptonic,
      semileptonic, hadronic, etc.
\item {\tt MODULE} {\tt cutvalues} -- contains values of kinematical cuts.
\item {\tt FUNCTION} {\tt ddilog} -- calculates
       the real part of a dilogarithm for real arguments.
\item {\tt MODULE} {\tt distribs} -- contains parameters of distributions.
\item {\tt MODULE} {\tt drivec} -- contains driving flags and some kinematical
      variables.
\item {\tt SUBROUTINE} {\tt eeee} -- returns a contraction of a quartic
      gauge boson coupling with four complex four vectors.
\item {\tt FUNCTION} {\tt eee} -- returns a contraction of a triple gauge
      boson coupling with three complex four vectors.
\item {\tt SUBROUTINE} {\tt eeff1} -- returns the squared matrix element
      averaged over initial spins and summed over final spins and colors of
      $\epm \ra f \bar{f}$.
\item {\tt SUBROUTINE} {\tt eeffffcc1a} -- returns the squared matrix element
      averaged over initial spins and summed over final spins and colors of
      (\ref{born}) for a charged current processes,  ${\tt iproc} < 11$
      or $ 14 \le {\tt iproc} < 21$, and
      the matrix element squared in the double $W$ resonance
      approximation in arbitrary linear gauge.
\item {\tt SUBROUTINE} {\tt eeffffcc1} -- returns the same as
      {\tt eeffffcc1a} but in the unitary gauge.
\item {\tt SUBROUTINE} {\tt eeffffcc2} -- returns the squared matrix element
      averaged over initial spins and summed over final spins and colors of
      (\ref{born}) for the mixed charged and neutral current processes,
      $ 11 \le {\tt iproc} < 14$. These matrix elements can be also obtained
      with {\tt eeffffnc3}.
\item {\tt SUBROUTINE} {\tt eeffffcc3} -- returns the same as
      {\tt eeffffcc2} for  $ \epm \ra \bar{\nu_e}\nu_e e^+e^-$,
      ${\tt iproc} = 21$.
\item {\tt SUBROUTINE} {\tt eeffffgid1} -- returns the squared matrix element
      averaged over initial spins and summed over final spins and colors of
      the neutral current processes (\ref{brems}) containing identical
      leptons, $ 401 \le {\tt iproc} < 406$.
\item {\tt SUBROUTINE} {\tt eeffffgid2} -- returns the same as
      {\tt eeffffgid1} for the neutral current processes (\ref{brems})
      containing identical quarks, $ 406 \le {\tt iproc} < 412$.
\item {\tt SUBROUTINE} {\tt eeffffgid3} -- returns the same as
      {\tt eeffffgid1} for the neutral current processes (\ref{brems})
      containing identical electron neutrinos, $ {\tt iproc} = 412$.
\item {\tt SUBROUTINE} {\tt eeffffgnc1} -- returns the squared matrix element
      averaged over initial spins and summed over final spins and colors of
      the neutral current processes (\ref{brems}) containing neither
      virtual gluons nor an external electron, $ 21 \le {\tt iproc} < 201$.
\item {\tt SUBROUTINE} {\tt eeffffgnc2} -- returns the same as
      {\tt eeffffgnc1} for the neutral current processes (\ref{brems}) not
      containing virtual gluons, $ 201 \le {\tt iproc} < 211$.
\item {\tt SUBROUTINE} {\tt eeffffgnc3} -- returns the same as
      {\tt eeffffgnc1} for the neutral current processes (\ref{brems})
      containing $\bar{\nu}_e\nu_e$-pair, $ 211 \le {\tt iproc} < 301$.
\item {\tt SUBROUTINE} {\tt eeffffgnc3} -- returns the same as
      {\tt eeffffgnc1} for the mixed charged and neutral current processes
      (\ref{brems}), $ 301 \le {\tt iproc} < 401$.
\item {\tt FUNCTION} {\tt eeffffgs} -- returns the soft bremsstrahlung
      correction of Eq.~(\ref{softint}) for any channel of (\ref{born}).
\item {\tt SUBROUTINE} {\tt eeffffid1} -- the same as
      {\tt eeffffgid1} for (\ref{born}).
\item {\tt SUBROUTINE} {\tt eeffffid2} -- the same as
      {\tt eeffffgid2} for (\ref{born}).
\item {\tt SUBROUTINE} {\tt eeffffid3} -- the same as
      {\tt eeffffgid3} for (\ref{born}).
\item {\tt SUBROUTINE} {\tt eeffffnc1} -- the same as
      {\tt eeffffgnc1} for (\ref{born}).
\item {\tt SUBROUTINE} {\tt eeffffnc2} -- the same as
      {\tt eeffffgnc2} for (\ref{born}).
\item {\tt SUBROUTINE} {\tt eeffffnc3} -- the same as
      {\tt eeffffgnc3} for (\ref{born}).
\item {\tt SUBROUTINE} {\tt eeffffnc4} -- the same as
      {\tt eeffffgnc4} for (\ref{born}).
\item {\tt SUBROUTINES} {\tt eev, eve} and {\tt vee} (contained in {\tt
      gwwv.f}) -- return contractions
      of a triple gauge boson coupling with two polarization vectors
      leaving, respectively, the third, second and first Lorentz index
      uncontracted.
\item {\tt SUBROUTINE} {\tt fefefef} -- returns matrix elements of the form
      $\bar{u}_1/\!\!\!\Gamma_1S_{F1}/\!\!\!\Gamma_2S_{F2}/\!\!\!\Gamma_3u_2$,
      where $\Gamma_{i}$, $i=1,2,3$, and $S_{Fj}$, $j=1,2$,
      denote the SM gauge boson-fermion coupling and Feynman propagator
      of a fermion, respectively.
\item {\tt SUBROUTINE} {\tt fefefefh} -- returns the same as {\tt fefefef}
      including contributions of the Higgs boson exchange.
\item {\tt SUBROUTINE} {\tt fefef} -- returns matrix elements of the form
      $\bar{u}_1/\!\!\!\Gamma_1S_{F1}/\!\!\!\Gamma_2u_2$.
\item {\tt SUBROUTINE} {\tt fefefh} -- returns the same as {\tt fefef}
      including contributions of the Higgs boson exchange.
\item {\tt SUBROUTINE} {\tt fef} -- returns matrix elements of the form
      $\bar{u}_1/\!\!\!\Gamma_1u_2$.
\item {\tt SUBROUTINE} {\tt fefh} -- returns the same as {\tt fef}
      including contributions of the Higgs boson exchange.
\item {\tt SUBROUTINE} {\tt ffvgnew} -- returns four vectors
      representing contributions resulting from an attachment
      of a photon to the SM gauge boson-fermion vertex including
      a possible Higgs boson exchange contribution.
\item {\tt SUBROUTINE} {\tt fhf} -- returns matrix elements
      $g_{Huu}\Delta_H\bar{u}u$, where $g_{Huu}$ is the Higgs boson
      coupling to a fermion represented by spinor $u$ and $\Delta_H$ is
      the Feynman propagator of the Higgs boson.
\item {\tt SUBROUTINE} {\tt fsfa} -- returns matrix elements
      $g_{Suu}\Delta_S(\xi)\bar{u}u$, where $g_{Su_1u_2}$ is the scalar boson
      coupling to fermions represented by spinors $u_1, u_2$ and
      $\Delta_S(\xi)$ is the Feynman propagator of a scalar boson in an 
      arbitrary linear gauge.
\item {\tt SUBROUTINE} {\tt fsf} -- returns the same as {\tt fsfa} in
      the unitary gauge.
\item {\tt SUBROUTINE} {\tt fvfa} -- returns a set of four vectors
      of the form $\bar{u}_1\gamma_{\nu}(g_V^{(-)}P_-  + g_V^{(+)}P_+)$
      $\times u_2 D_V^{\nu\mu}(\xi)$, where $u_1$, $u_2$ are fermion spinors,
      $P_{\pm}=(1 \pm \gamma_5)/2$, are chirality projectors, $g_V^{(\pm)}$ are
      the gauge boson-fermion couplings of definite chirality and
      $D_V^{\nu\mu}(\xi)$ is the propagator of a gauge boson $V$ in arbitrary
      linear gauge.
\item {\tt SUBROUTINE} {\tt fvf} -- returns the same as {\tt fvfa} in
      the unitary gauge.
\item {\tt MODULE} {\tt gaugepar} -- contains driving flags and gauge
      parameters for calculations in arbitrary linear gauge.
\item {\tt MODULE} {\tt genpv12} -- contains matrix elements representing
      the photon and $Z$ boson coupling to initial state fermions.
\item {\tt MODULE} {\tt genpv12g} -- contains matrix elements representing
      the photon and $Z$ boson coupling to initial state fermions including
      photon radiation from initial lines.
\item {\tt MODULE} {\tt helproc} -- contains numbers of possible helicity
      states of external particles.
\item {\tt SUBROUTINE} {\tt inipart} -- calculates content of modules
      {\tt genpv12} and {\tt genpv12g}.
\item {\tt MODULE} {\tt inprms} -- contains initial input parameters.
\item {\tt FUNCTION} {\tt iomega} -- calculates the soft bremsstrahlung
      phase space integrals (\ref{iomega}) and (\ref{iomegaii}).
\item {\tt SUBROUTINE} {\tt kineeff1} -- returns the four momenta, phase
      space normalization and flux factor for a $2 \ra 2$ process in the CMS.
      The phase space is parametrized according to Eq.~(\ref{dps}).
\item {\tt SUBROUTINE} {\tt kineeffff} -- returns the four momenta, phase
      space normalization and flux factor for a 2 $\ra$ 4 process in
      the CMS. The phase space is parametrized according to Eq.~(\ref{dpsb1})
      or (\ref{dpsb2}), dependent on selected options.
\item {\tt SUBROUTINE} {\tt kineeffffg} -- returns the same as
      {\tt kineeffff} for a 2 $\ra$ 5 process in
      the CMS. The phase space is parametrized according to Eq.~(\ref{dps1}),
      (\ref{dps2}) or (\ref{dps3}), dependent on selected options.
\item {\tt SUBROUTINE} {\tt kineetbff} -- returns the same as
      {\tt kineeffff} with the phase space parametrized according to
      Eq.~(\ref{dpsb2}).
\item {\tt SUBROUTINE} {\tt kineetbffg} -- returns the same as
      {\tt kineeffff} with the phase space parametrized according to
      Eq.~(\ref{dps2}).
\item {\tt SUBROUTINE} {\tt kinff} -- returns the final state four momenta
      and phase space normalization for a $2 \ra 2$ process in the CMS.
\item {\tt SUBROUTINE} {\tt kinffg} -- returns the same as {\tt kinff}
      for a $2 \ra 3$ process in the CMS.
\item {\tt FUNCTION} {\tt lamsq} -- the kinematic lambda function,
      $\lambda(\sqrt{x},\sqrt{y},\sqrt{z})$.
\item {\tt SUBROUTINE} {\tt mateeee} -- returns polarized matrix elements
      of $\epm \ra \emp\emp$.
\item {\tt SUBROUTINE} {\tt mateeeeg} -- returns polarized matrix elements
      of $\epm \ra \emp\emp\gamma$.
\item {\tt MODULE} {\tt mathprms} -- contains necessary arithmetical constants.
\item {\tt SUBROUTINE} {\tt matllll} -- returns polarized matrix elements
      of $\epm \ra l \bar{l} l \bar{l}$.
\item {\tt SUBROUTINE} {\tt matllllg} -- returns polarized matrix elements
      of $\epm \ra l \bar{l} l \bar{l} \gamma$.
\item {\tt SUBROUTINE} {\tt matnnnn} -- returns polarized matrix elements
      of $\epm \ra \nu_e \bar{\nu}_e\nu_e \bar{\nu}_e$.
\item {\tt SUBROUTINE} {\tt matnnnng} -- returns polarized matrix elements
      of $\epm \ra \nu_e \bar{\nu}_e\nu_e \bar{\nu}_e\gamma$.
\item {\tt SUBROUTINE} {\tt matqqqq} -- returns polarized matrix elements
      of $\epm \ra q \bar{q} q \bar{q}$.
\item {\tt SUBROUTINE} {\tt matqqqqg} -- returns polarized matrix elements
      of $\epm \ra q \bar{q} q \bar{q}\gamma$.
\item {\tt MODULE} {\tt parproc} -- contains parameters of a specific process.
\item {\tt MODULE} {\tt parsm} -- contains SM couplings and squared masses.
\item {\tt SUBROUTINE} {\tt parspec} -- returns parameters of a specific
      process (\ref{born}) and (\ref{brems}), calculates the top quark
      and Higgs boson
      width to the lowest order of SM, the branching ratios for
      the narrow width approximations; initializes weights for the
      Monte Carlo integration.
\item {\tt SUBROUTINE} {\tt psneeffff} -- returns a phase space normalization
      as that of {\tt kineeffff} for a given set of external particle
      momenta.
\item {\tt SUBROUTINE} {\tt psneeffffg} -- returns a phase space normalization
      as that of {\tt kineeffffg} for a given set of external particle
      momenta.
\item {\tt SUBROUTINE} {\tt psneetbff} -- returns a phase space normalization
      as that of {\tt kineetbff} for a given set of external particle
      momenta.
\item {\tt SUBROUTINE} {\tt psneetbffg} -- returns a phase space normalization
      as that of {\tt kineetbffg} for a given set of external particle
      momenta.
\item {\tt SUBROUTINE} {\tt recpol} -- returns real polarization vectors of
      a vector boson in the rectangular basis.
\item {\tt FUNCTIONS} {\tt scc, src} and {\tt srr} (contained in
      {\tt dotprod.f}) -- return the Minkowski dot product of two complex,
      real and complex, and two real four vectors.
\item {\tt SUBROUTINE} {\tt spheric} -- returns spherical components {\tt ps}
       of a four vector $p^{\mu}$, {\tt ps}$=\\
       (|{\bf p}|,\cos{\theta},
       \sin{\theta},\cos{\phi},\sin{\phi})$, with $\theta$ and $\phi$ being
       a polar and azimuthal angles of momentum {\bf p}.
\item {\tt SUBROUTINE} {\tt spinc} -- returns the contractions:
      $p^0 I - {\bf p \cdot \sigma}$ and $p^0 I + {\bf p \cdot \sigma}$,
      where $p^{\mu}=(p^0,{\bf p})$ is a complex four vector, $I$ is the
      $2 \times 2$ unit matrix and $\sigma$ are the Pauli matrices.
\item {\tt SUBROUTINE} {\tt spinornew} -- returns helicity spinors in the Weyl
      representation.
\item {\tt SUBROUTINE} {\tt spinr} -- returns the same contractions as
      in {\tt spinc} for a real four vector $p^{\mu}$.
\item {\tt MODULE} {\tt topapprox} -- contains SM lowest order widths of
      the top quark in different approximations.
\item {\tt SUBROUTINE} {\tt uep} -- returns a set of spinors of the form
      $\bar{u} /\!\!\!\varepsilon(g_V^{(-)}P_-  + g_V^{(+)}P_+) S_F$,
      where $S_F$ is a Feynman propagator of an internal fermion and the
      remaining notation is the same as in {\tt fvfa}.
\item {\tt SUBROUTINE} {\tt usp} -- returns a set of spinors:
      $\bar{u} (g_S^{(-)}P_-  + g_S^{(+)}P_+) S_F$,
      where $S_F$ is a Feynman propagator of an internal fermion and
      $g_S^{\pm}$ are complex chirality couplings of the scalar boson to
      the fermion represented by spinor $u$.
\item {\tt SUBROUTINE} {\tt wwidth} -- returns the partial width of the
      $W$-boson averaged over initial spins and summed over final spins,
      and colors for hadronic decay modes.
\end{itemize}

\subsection{Run output}
A sample of the listing of the test run output is given in an attached 
file {\tt test0}.
It contains a specification of the process, information on the scheme choice,
and values of the relevant physical parameters in the very beginning.
If cuts have been imposed ({\tt icut = 1}) then there comes information about
specified cuts.
Then, if {\tt iscan = 1}, results of the initial scan and resulting 
weights for the actual run are printed.
If calculated weights do not add up exactly
to 1 the last weight is changed slightly in order to satisfy the normalization
condition. This change is completely irrelevant numerically,
but the message about it is printed.
The final result for the total cross
section is called {\tt Integral}. Its value is printed together
with the standard deviation and the actual number of calls used in the
calculation. 
Finally, if {\tt imc = 1} then
information on events acceptance efficiency is given, that means a fraction
of accepted weight 1 events.
The same kind of output is printed for the soft (without initial scan)
and hard bremsstrahlung for each value of the CMS energy.
In the very end, all the calculated total cross sections together with
the corresponding standard deviations are collected in the tabular form.

If the event generation
option is switched on, i.e. {\tt imc = 1}, then the corresponding
number of unweighted events will be printed out as collections of final
state particle four momenta in a separate file called {\tt events.dat}.
Whenever a maximum value of the cross section initially assigned in
{\tt ee4fg.f}, or found in the result of the initial scan is overflown
a corresponding message informing about it is printed in {\tt events.dat}.
The result of integration is still valid, however, if the program is run as
an event generator, it should be rerun. How to proceed in this situation will
be described in the next section.

\section{Use of the program}
Up to now the program have been run only on Unix/Linux platforms.
In order to run the program, the user should select
a specific name of a {\tt FORTRAN 90/95} compiler, desired options and the
name of the output file in the {\tt makefile}.
The output file is called {\tt test} at present.
The program can be then compiled, linked and run by executing a single
command\\[2mm]
{\tt make test}.\\[2mm]
The results of the test run should reproduce those contained
in files {\tt test0} and {\tt events.dat0}.

The program can be run as the MC event generator of unweighted
events by selecting \\[2mm]
{\tt imc = 1}\\[2mm]
in {\tt ee4fg.f}. It is then recommended to run the program for a single CMS
energy and to perform a scan with a relatively large number of calls,
{\tt nscan0} or {\tt nscanh}. Attention should be paid to possible messages
informing about updates of the maximum weight. In this case the program
should be rerun, however, this time without the initial scan. The initial
integration weights {\tt aw0} or {\tt aw} in {\tt parspec.f}
and the maximum weight {\tt crmax}
in {\tt ee4fg.f} should be updated according to the results of the
prior scan. At present the efficiency of events acceptance is relatively
low. However, in view of a relatively fast performance of the program,
this should not be a serious limitation for the user.
%
%


%

\end{document}